\documentclass[twocolumn,showpacs,preprintnumbers,amsmath,amssymb,prl,
superscriptaddress]{revtex4-1}

\usepackage{graphicx}
\usepackage{color}
\usepackage{epsfig}
\usepackage{overpic}
\usepackage{ulem}

\newcommand{\be}{\begin{eqnarray}}
\newcommand{\ee}{\end{eqnarray}}

\begin{document}
 
\title{BCS theory of driven superconductivity}
\author{Andreas Komnik}
\affiliation{Institut f\"ur Theoretische Physik, Universit\"at Heidelberg,
Philosophenweg 12, D-69120 Heidelberg, Germany}
\author{Michael Thorwart}
\affiliation{I.\ Institut f\"ur Theoretische Physik,
Universit\"at Hamburg,
Jungiusstr.\ 9,
20355 Hamburg, Germany}
\affiliation{The Hamburg Center for Ultrafast Imaging, Luruper Chaussee 149,
22761 Hamburg, Germany}

\date{\today}

\begin{abstract}
We study the impact of a time-dependent external driving of the lattice phonons
in a minimal model of a BCS superconductor. Upon evaluating the driving-induced
vertex corrections of the phonon-mediated electron-electron interaction, we
show that parametric phonon driving can be used to elevate the critical
temperature $T_c$, while a dipolar phonon drive has no effect. We provide simple
analytic expressions for the enhancement factor of $T_c$.
Furthermore, a mean-field analysis of a nonlinear phonon-phonon interaction
also shows that phonon anharmonicities further amplify $T_c$.
Our results hold universally for the large class of normal BCS superconductors. 
\end{abstract}


\maketitle

Quantum many-body systems which are driven far away from thermal
equilibrium represent an increasingly fascinating realm of condensed matter
physics, since recent progress in the experimental techniques has made it
possible to manipulate condensed matter quantum states by strong external
fields \cite{CavalleriReview2015}. Light can strongly modify phases of 
correlated quantum many-body systems. For instance, strong time-dependent
fields can induce transient superconducting phases in different material
classes \cite{fausti_light-induced_2011,mankowsky_nonlinear_2014,
hu_optically_2014,kaiser_optically_2014,foerst_2015,singla_2015,
mitrano_optically_2015}. Moreover, electromagnetic irradiation
 can induce a collapse of 
long-range ordered charge-density wave phases
\cite{schmitt_transient_2008,yusupov_coherent_2010,hellmann_ultrafast_2010,
rohwer_collapse_2011}, deconstruct insulating phases
\cite{rini_control_2007,hilton_enhanced_2007,
liu_terahertz-field-induced_2012-1}, or break up Cooper pair quasiparticles
\cite{graf_nodal_2011,smallwood_tracking_2012,matsunaga_higgs_2013}. 

Conceptual insight into the possible physical mechanisms has been greatly
advanced recently
\cite{FoerstNatPhys2011,Subedi2014,Kemper2015,Raines2015,Sentef2015,
Murakami2015,Demler2015}. In the presence of strong lattice anharmonicities, 
 the nonlinear coupling of a resonantly driven phonon to other Raman-active 
modes leads to a rectification of a directly excited infrared-active mode and to
a net displacement of the crystal along the coordinate of all anharmonically
coupled modes \cite{FoerstNatPhys2011,Subedi2014}. Selective vibrational
excitation can
also drive high-$T_C$ cuprates into a transiently enhanced
superconducting state. Moreover, on the basis of the non-equilibrium Keldysh
formalism, partial melting of the superconducting phase by the pump field has
been identified \cite{Kemper2015}. Furthermore, an advanced extension of the
single-layer $t$-$J$-$V$ model of cuprates
to three dimensions has been used to show that an optical pump can be
used to suppress the charge order and enhance superconductivity
\cite{Raines2015}. In an effective approach on the basis of a driving-induced
reduction of the electronic hopping amplitude, the resulting increase of the
density of states near the Fermi edge has been shown to enhance
superconductivity \cite{Sentef2015}. Using the nonequilibrium dynamical
mean-field theory for a strongly coupled
electron-phonon system, a strong electron-mediated phonon-phonon interaction
has been revealed \cite{Murakami2015}. These theoretical approaches are all 
very advanced and specialized to particular classes of systems and are rather
successful in explaining experimental data for specific materials. Yet, it is
still 
desirable to establish and analyze minimal models to reveal the
fundamental mechanisms in terms of simple and elegant analytical results. 
Very recently, such a minimal model of a strongly
driven electron-phonon Hamiltonian has been analyzed upon using Floquet
formalism \cite{Demler2015}. A Floquet BCS gap equation is derived which calls
for a numerical solution and does not permit closed analytic results. 

In this work, we aim to obtain a rather general and explicit analytical result 
to illustrate the driving-induced elevation of the critical temperature of a
normal superconductor by extending the conventional BCS theory.  We go beyond
the conventional approaches, which usually consider the modification of the
distribution function of charge carriers (see e.~g. \cite{Eliashberg}) and
 consider the standard Fr\"ohlich-type electron-phonon Hamiltonian
with linear phonons subject to a time-dependent external driving. 
We show that a simple
dipolar coupling of the driving field to the phonon displacement coordinates
only yields a scalar phase shift and does not modify the 
electron-phonon interaction vertex. In contrast to that, a parametric driving of the
phonon frequencies strongly modifies the retarded Green's function,
 thereby changing the effective electron-electron attraction in a fundamental way. 
 In order to quantify these effects, we introduce an elevation factor
$\eta$ of the  critical temperature which directly can be calculated in 
our approach. In the limits of weak and
strong driving fields, we obtain simple expressions for $\eta$, which show how
the critical temperature can be enhanced even if the driving is nonresonant. 
Finally, we show that a parametric phonon drive combined with a phonon-phonon
interaction can induce an additional elevation of the critical temperature. 
This is apparent already on the level of a mean-field treatment of the nonlinear phononics. 

{\it Minimal model of a driven BCS superconductor -- }
The canonical modeling of the superconducting materials is based on the
Fr\"ohlich-type Hamiltonian ($\hbar=1$)
\be            \label{Eq0}
 H_0 = H_\psi[c_{\bf k}] + H_\Omega + \frac{\lambda}{\sqrt{\cal V}} \sum_{{\bf
k, q}} (a^\dag_{\bf q} - a_{-{\bf q}}) \, c^\dag_{\bf k - q} \, c_{\bf k} \, ,
\ee
where $H_\psi[c_{\bf k}] =  \sum_{\bf k} (\epsilon_{\bf k} - \mu) \, c^\dag_{\bf
k } \, c_{\bf k}$ is a Hamiltonian of the electronic conductance band, 
\be            \label{Eq1}
 H_\Omega =\sum_{\bf q} \Omega_{\bf q}  a_{\bf q}^\dag a_{\bf q} 
\ee
describes the phonon degrees of freedom with the dispersion $
\Omega_{\bf q} $ and 
$\lambda$ is the electron-phonon interaction strength. ${\cal V}$ is the volume
of 
the sample. The deflection field 
$Q_{\bf q} = a^\dag_{\bf q} + a_{\bf -q}$ of the phonons is the
Fourier transform of the phonon coordinate. The BCS theory is build upon
the fact that $Q_{\bf q}$ can be integrated over, such that an exact 
effective action 
\be
 S &=& S_0 + \lambda^2  \sum_{\bf q, k, k'} \int d t d t' \, 
 c^\dag_{\bf k +q } (t) \, c_{\bf k} (t) \, G({\bf q}, t-t') \, 
 \nonumber \\
 &\times& c^\dag_{\bf k'-q } (t') \, c_{\bf k'} (t') \, 
\ee
results. Here, $G({\bf q}, t-t')$ is the Green's function (GF) of the
deflection field $Q_{\bf q}$. Its retarded component is canonically defined as  
\be   \label{GRdef}
 && G^R({\bf q, q'}, t-t') =
  - i \Theta(t - t') 
 \nonumber \\
 &\times& \langle Q_{\bf q}(t) \, Q_{\bf q'}(t') - Q_{\bf q'}(t') \, Q_{\bf
q}(t) \rangle \, .
\ee
It generates an approximative interaction vertex amplitude $V({\bf q},\omega)$ of
an effective electron-electron interaction mediated by the phonons and is at
the heart of BCS theory of superconductivity. 
To lowest order in $\lambda$, one then obtains for the interaction vertex 
\be  \label{nakedVertex}
 V({\bf q}, \omega) = G_0^R({\bf q, -q}, \omega)=\frac{2 \Omega_{\bf
q}}{\omega^2 - \Omega^2_{\bf q}} \, ,
\ee
where $\omega$ is the energy transfer during the  scattering of the electron
pair.  Obviously, if $\omega^2 < \Omega^2_{\bf q}$,  the effective
interaction is attractive, thus leading to the Cooper instability and
superconducting ground state \cite{Abrikosov1975}.   
In general, the larger the overall scale of $\Omega_{\bf q} $ is, the larger
is the range of energies $\omega$ and the higher is the number of electrons, for
which the mutual interaction becomes attractive. This is accompanied by an
increase of the critical temperature $T_c$, at which the superconducting gap
vanishes.  

The critical temperature in a BCS superconductor in a simplest model of an
attractive constant potential of strength $V_0$ is given by ($k_B=1$) 
\be    \label{Tc}
 T_c \simeq \omega_D e^{- 1/[V_0 \rho(E_F)]} \, , 
\ee
where $\rho(E_F)$ is the density of the electronic states at the Fermi edge and
$\omega_D$ is the Debye frequency which fixes the characteristic energy scale
for the phonon degrees of freedom.  
The expression in Eq.\ \eqref{Tc} can be considered as generic if one 
interprets $V_0$ as an effective parameter which measures the strength of the
(in general energy and momentum dependent) attractive potential. 
There are basically three different options to increase $T_c$ by changing one of
the above parameters. We shall consider two of them: (i) the enhancement of the
effective attraction $V_0$, and, (ii) the increase of $\omega_D$. 

One way to modify the denominator of Eq.\ \eqref{GRdef} is to drive the phonons
by strong electromagnetic external THz fields. The driving can induce 
phonon excitations in sequential steps, in which the phonons are directly
excited by applied EM field pulses. Alternatively, infrared-active phonon modes
with a finite dipole moment can be excited, and due to nonlinear phonon
coupling, normal phonon Raman modes of the crystal are excited
\cite{FoerstNatPhys2011,Demler2015}. We choose not to concentrate on these
intricacies as they are strongly material-dependent and thus nonuniversal and
consider the driving as acting directly to the relevant phonon mode. There
are essentially two qualitatively different possibilities, the dipolar (or
linear) driving where the drive couples to the phonon deflection field, and the
parametric (or quadratic) driving where the drive modulates the phonon
frequencies. 

{\it Dipolar phonon driving -- } The dipolar phonon driving by an explicitly
time-dependent driving field $\Delta_{\bf q}(t)$ does not influence the retarded
GF $G^R({\bf q, q'}, \omega)$. This immediately follows when we 
 replace  Eq.\ \eqref{Eq1} by   
 $ H_\Omega =\sum_{\bf q} \Omega_{\bf q}  a_{\bf q}^\dag a_{\bf q}  +
\Delta_{\bf q}(t) (a_{\bf q}^\dag - a_{\bf -q}) 
$.
The electron-phonon coupling strength $\lambda$ is quite weak in most of the
known superconducting materials. For this reason, the leading behavior of the
$G^R({\bf q, q'},\omega)$ is dominated by the  contribution of the
phonon subsystem only. Solving the equations of motion for $H_\Omega$, one
readily finds  
$a_{\bf q}(t) = [a_{\bf q}(0) + f(t)]\, e^{- i  \Omega_{\bf q}  t}$, 
where 
 $f(t) = - i \int^t d t' \, \Delta_{\bf q}(t') e^{i \Omega_{\bf q}  t'}$  
is a simple time-dependent scalar phase shift. As the retarded GF is a
commutator
of fields, a mere shift of them does not affect the GF at all. Thus, we conclude
that within our approximation the linear driving does not affect the
conventional BCS superconductivity picture. 

{\it Parametric phonon driving -- } The parametric driving enters via 
the Hamiltonian 
\be   \label{HOmega}
  H_\Omega =\sum_{\bf q}[ \Omega_{\bf q}  + \Delta_{\bf q}(t) ] \, a_{\bf
q}^\dag a_{\bf q}   \, . 
\ee
It is, e.g., realized indirectly by resonantly driving infrared-active phonon
modes with a finite dipole moment, which couple quadratically to normal Raman
modes of the crystal
\cite{mankowsky_nonlinear_2014,FoerstNatPhys2011,Subedi2014,CavalleriReview2015}
or by 
the quadrupole component of an electromagnetic field. 
The trivial case is the static driving, i.e.,  $\Delta_{\bf q}(t) = \Delta_{\bf
q}$, which simply is an increase of phonon frequencies. As the Debye
frequency rises as well, an increase of $T_c$ is obvious. This effect
is known and is experimentally detected in crystals subject to high pressure. 
In the dynamical case, the solution for the time evolution equation is obviously
$a_{\bf q}(t) = a_{\bf q}(0) \, e^{- i  \alpha(t)}$ with the phase 
$\alpha(t) = \Omega_{\bf q}  t + \int_0^t d t' \, \Delta_{\bf q}(t')$. Then, 
the retarded GF of Eq.\ \eqref{GRdef} follows as 
\be \label{GRalpha}
 G^R({\bf q, q'},t,t') = - i \delta_{\bf -q, q'}  \Theta(t - t') 
 \nonumber \\ \times
  \sum_{\pm} (\pm1) e^{ \pm i
[\alpha(t)-\alpha(t')]}\, .
\ee
Without restricting the generality, we henceforth assume periodic
time-dependent driving in the form  
$  \Delta_{\bf q}(t) =  \Delta_{\bf q} \cos (\Gamma_{\bf q} t)$,
where $ \Delta_{\bf q}$ is the strength and $\Gamma_{\bf q}$ is the
frequency of the driving. 
%
After a Fourier expansion with respect to
the time difference $t-t'$, we obtain a result in terms of the $n$-th ordinary
Bessel
function \cite{gradshteyn2007}:
\be              \label{fullGR}
&& G^R({\bf q, q'},\omega) =  i \delta_{\bf -q, q'} 
 \left\{
   \frac{- i \, 2  \Omega_{\bf q} }{\omega^2 -  \Omega_{\bf q}^2} 
   - i \sum_{n=1}^\infty
   \right. \\
 &&   \left. \frac{(\omega - \Omega_{\bf q}) J_n \left[ \frac{n y
}{\omega - \Omega_{\bf q}} \right]}{(\omega -
\Omega_{\bf q})^2-(n\Gamma_{\bf q}/2)^2} 
\nonumber
   -
    \frac{(\omega + \Omega_{\bf q}) J_n \left[ -\frac{ ny }{\omega +
\Omega_{\bf q} }\right]}{(\omega + 
\Omega_{\bf q})^2-(n\Gamma_{\bf q}/2)^2} 
\right\} \, , 
\ee
where 
\be     \label{pdef}
y = \Delta_{\bf q} \sin \left[ \Gamma_{\bf q} \frac{(t+t')}{2} \right]  \, 
\ee
explicitly depends on the evolution time $\tau = ( t+t')/2$.
We recover the zero-order contribution of Eq.\ \eqref{nakedVertex} as 
the first term of the r.h.s.\ of Eq.\ \eqref{fullGR}. Moreover, the multiphonon 
parametric resonances are apparent from the denominators when $2
\tilde{\omega}_{\bf q} = n \Gamma_{\bf q}$. 

To proceed, we exploit that the typical driving frequency in the experiments is
in the THz regime, which is slightly smaller than the
typical Debye frequency of superconducting materials. Hence, we may average
over the period of the external driving with respect to $\tau$. For the
time-averaged Bessel functions, we then obtain
\be
 && \bar{J}_n \left[ \frac{n y }{\omega - \Omega_{\bf
q}} \right] =
 \frac{1}{2} \int_{-1}^1 d x \, J_n \left[ \frac{x n \Delta_{\bf
q}}{\omega - \Omega_{\bf q}} \right]
\\  \nonumber 
 &=& \frac{1}{(n+1)!} \left[ \frac{n \Delta_{\bf q}}{2 (\omega - \Omega_{\bf
q})}\right]^n
 \nonumber \\ &\times&
 {}_1\!F_2 \left[ \frac{1+n}{2}; \frac{3+n}{2}, 1+n; 
  -   \left( \frac{n \Delta_{\bf q}}{ 2(\omega - \Omega_{\bf q})}\right)^2
\right]
\ee
for even $n$ and zero otherwise. Here, ${}_1\!F_2$ denotes the
hypergeometric function \cite{gradshteyn2007}.  
Its maximum is of the order of $1$ for $n<2$ for any 
argument and it decays exponentially for $n>2$. Hence, we may focus 
 on the lowest order term $n=2$ only. The physical meaning is immediate. $n$
denotes the number of phonons which participate in the
renormalization of the GF by the vertex. The odd phonon numbers do not
contribute for symmetry reasons. The larger $n$, the more efficient is the
mutual cancellation during averaging. As a result, only the two phonon
process survives which is the parametric resonance. 

In order to quantify the enhancement of the interaction vertex around the Fermi
edge, we define the enhancement factor $\eta= G^R({\bf q, q'},
0)/G_0^R({\bf q, q'}, 0)$ as the ratio of the two retarded GF 
in the low-energy limit. Moreover, we may
exploit the asymptotic behavior of the hypergeometric function for $n=2$ for $x
\ll 1$ in the form { $\frac{1}{6} x^2  {}_1\!F_2 [3/2; 5/2, 3; - x^2] = x^2/6 +
O(x^4)$} to assess the quantitative behavior of the GF in Eq.\ \eqref{GRdef} in
the vicinity of the Fermi edge $\omega \to 0$. Hence, for weak driving
$\Delta_{\bf q} \ll \Omega_{\bf q}$, we obtain 
\be
\eta= \frac{G^R({\bf q, q'}, 0)}{G_0^R({\bf q, q'}, 0)} = 1 +
\frac{\Delta^2_{\bf q}}{6 ( \Omega^2_{\bf q} -  \Gamma^2_{\bf q})} \, .
\ee
$\eta > 1$ implies a relative enhancement of the attractive interaction around
the
Fermi edge and thus an increase in $T_c$, since $\eta$ enters in the expression
for the critical temperature as a factor renormalizing the electron-phonon
coupling strength according to $V_0 \to \eta V_0$. This occurs for subresonant
driving $\Omega_{\bf q} > \Gamma_{\bf q}$, which is the most realistic regime
from the point of view of contemporary experiments, and can, at least in
principle, become quite large. In the opposite case of superresonant driving
$\Omega_{\bf q} < \Gamma_{\bf q}$, there is a decrease of $T_c$. 
 This kind of transition should be experimentally observable. 

In the limit of strong driving $\Delta_{\bf q}
\gg \Omega_{\bf q}$, we obtain with ${\frac{1}{6} x^2  {}_1\!F_2 [3/2; 5/2, 3; -
x^2] = 1/(2|x|) + O(1/x^2)}$ for $x \gg 1 $ the enhancement factor 
\be
 \eta = 1 + \frac{\Omega^3_{\bf q}}{ 2 \Delta_{\bf q} ( \Omega^2_{\bf q} - 
\Gamma^2_{\bf q})} \, .
\ee
It shows the similar dependence of $\Omega_{\bf q}$ and $\Gamma_{\bf q}$.
Although an estimate of the validity region of our approximation is more
involved, we believe our results to hold for $\eta \sim 1-2$. 

{\it Nonlinear phononics -- } Next, we address the role of the 
phonon anharmonicity. On the
microscopic level, it arises due to a nonlinear interaction between the
 phonons. Usually, one encounters two different kinds: three-
and four-phonon interaction processes. They are described by the Hamiltonians  
\be
 H_3 &=&  \sum_{\bf q, k}  M_3({\bf q, k}) Q_{\bf k} \, Q_{\bf q} \, Q_{\bf - k
- q} \, , \\
  H_4 &=& \sum_{\bf q, k, p}  M_4({\bf q, k, p})  Q_{\bf k} \, Q_{\bf q} \,
Q_{\bf - k - p} \, Q_{\bf - q + p} \, , 
\ee
where $M_{3,4}$ are the corresponding interaction amplitudes. 
As a rule,  they are small and the appropriate way to 
assess their influence is the perturbation
theory. It turns out that the three-phonon self-energy vanishes exactly for 
homogeneous systems and is strongly suppressed in lattices with high symmetry
groups. Hence, we focus on the four-phonon process. We are interested in
the effective properties of one single phonon mode. Therefore, the most
important contribution is expected to be given by the nondiffractive scattering
processes of the given phonon mode on itself, when ${\bf p}=0$ and ${\bf k} =
{\bf q}$. The underlying effective Hamiltonian \cite{EnakiGolun2012} can be
inferred from the above one and one finds  
\be            \label{Eq10}
 H_\Omega =\sum_{\bf q} \Omega_{\bf q}  a_{\bf q}^\dag a_{\bf q} + \chi({\bf q})
\, a_{\bf q}^\dag a_{\bf q}^\dag a_{\bf q} a_{\bf q} \, .
\ee
The anharmonicity coefficient $\chi({\bf q})$ can be obtained from
$M_4({\bf q, q, 0})$ and is expected to be small. Since phonons at rest
do not exist we can write $\chi({\bf q}) \approx \chi_1 q$, where $q=|{\bf q}|$.
Although in a superconducting material at low temperatures, the
phonon expectation value $\langle a_{\bf
q}^\dag a_{\bf q} \rangle = N_{\bf q}$ is strongly suppressed, this is not the
case in presence of an external drive. Invoking a mean field
approximation, the effective Hamiltonian  is found
to be 
\be
 H_\Omega \approx \sum_{\bf q} \Omega_{\bf q}  a_{\bf q}^\dag a_{\bf q} +
\chi({\bf q}) \, a_{\bf q}^\dag N_{\bf q}  a_{\bf q}
 \nonumber \\
=  \sum_{\bf q} [\Omega_{\bf q}  + \chi({\bf q}) \, N_{\bf q} ] a_{\bf q}^\dag
a_{\bf q}  \, .
\ee
In equilibrium and without external driving, $N_{\bf q}$ is determined from the
self-consistency
condition $N_{\bf q} = \left[ e^{\beta (\Omega_{\bf q} + \chi_{\bf q} N_{\bf
q})} - 1\right]^{-1}$ and turns out to be smaller in comparison to the linear
system with $\chi_{\bf q}=0$. For this reason, the impact of the
anharmonic phonon subsystem on the electronic properties is negligible 
and does not induce any appreciable change in $T_c$ without driving 
\cite{FreericksMahan1996}. This
is completely different in a strongly driven system, where the phonon
population $N_{\bf q}$ is determined by the irradiation field. In this case, the
phonon subsystem is stiffer and is characterized by an effectively enhanced
Debye frequency $\omega_D$. In order to illustrate this feature, we consider 
the simplest case $\Omega_{\bf q}  = v_s q$, where $v_s$ is the bare sound
velocity of the crystal. Then,  $\Omega_\text{eff}({\bf q}) \equiv \Omega_{\bf
q} + \chi({\bf q}) \,
N_{\bf q}  \approx (v_s + \chi_1 N_{\bf q}) q$. Hence, the critical temperature
is renormalized according to $T_c \to \xi T_c$ with $\xi =
 1 + N_{\bf q} \chi_1/v_s$ and is thus increased.  

Hence, if a nonlinear superconducting material is exposed to
strong external parametric driving, the critical temperature can be increased
by two effects, so that Eq.~\eqref{Tc} is modified to
\be    \label{Tcnew}
  T_c \simeq \xi \, \omega_D \, e^{- 1/[\eta V_0 \rho(E_F)]} \, , 
\ee
when $\eta,\xi>1$. Overall, the theory is expected to hold quantitatively up to 
$\Delta_{\bf q}/\Gamma_{\bf q} \simeq 1$. It is important to realize that the
enhancement factor enters in the exponent of $T_c$.

{\it Conclusions -- } By considering a minimal model of a Fr\"ohlich-type BCS
Hamiltonian of a normal superconductor in presence of a time-dependent periodic
electromagnetic driving of the phonons, we illustrate the basic physical
mechanisms by which the critical temperature $T_c$ can be elevated.  We show
that
while a dipole (linear) driving cannot change 
$T_c$ of the material, quadratic (parametric) driving can enhance the
effective attractive phonon-mediated electron-electron interaction and thus
increase the critical temperature. The effect in this minimal model is
illustrated in terms of the enhancement factor of the interaction vertex caused
by the external driving. In the limits of weak and strong external phonon
driving, we find simple analytic results for the vertex enhancement. 
Furthermore, although an additional phonon anharmonicity does not change $T_c$
in BCS superconductors held at equilibrium, nonlinear phononics can provide an
additional contribution to the elevation of $T_c$, in that the effective Debye
frequency is renormalized. Finally, we note that the external phonon
drive also increases electron scattering, which in general suppresses Cooper
pairing.  Yet, it has been shown recently \cite{Demler2015} that the dynamic
enhancement of the formation of Cooper pairs addressed here dominates
over the increase of the scattering rate. These results in terms of a minimal
model shed new light on the essential ingredients needed for manipulating the 
characteristics of a BCS superconductor. A detailed analysis of the
quasiparticle decay processes and their interplay with enhanced interaction
vertex is an obvious avenue for further research \cite{KT2}.

We thank Reinhold Egger and Hermann Grabert for the careful reading of our
manuscript. 
AK is supported by the Heisenberg Program of the Deutsche
Forschungsgemeinschaft (Germany) under Grant No. KO 2235/5-1. MT acknowledges 
support from ``The Hamburg Center for Ultrafast Imaging'' funded by the DFG.


\end{document}